\documentclass[aps,twocolumn,superscriptaddress]{revtex4-1}
\usepackage[english]{babel}
\usepackage{graphicx,dcolumn,bm,amssymb,amsmath,amsfonts,amsthm,latexsym}
\usepackage{float,subfig,slashed,hyperref}
\hypersetup{
  linktocpage  = true,
  colorlinks   = true,
  urlcolor     = red,
  linkcolor    = black,
  citecolor    = blue
}

\begin{document}

\title{Comparison between the continuum threshold and the Polyakov loop as deconfinement order parameters}

\author{J.P.~Carlomagno}
\affiliation{IFLP, CONICET $-$ Dpto.\ de F\'{\i}sica, Universidad Nacional de La Plata, C.C. 67, 1900 La Plata, Argentina}
\affiliation{CONICET, Rivadavia 1917, 1033 Buenos Aires, Argentina}
\author{M.~Loewe}
\affiliation{Instituto de F\'isica, Pontificia Universidad Cat\'olica de Chile, Casilla 306, San\-tia\-go 22, Chile}
\affiliation{Centre for Theoretical and Mathematical Physics and Department of Physics,
University of Cape Town, Rondebosch 7700, South Africa}
\affiliation{Centro Cient\'{\i}fico-Tecnol\'ogico de Valpara\'{\i}so, Casilla 110-V, Valpara\'{\i}so, Chile}     


\begin{abstract}
We study the relation between the continuum threshold as function of the temperature $s_0(T)$ within finite energy sum rules and the trace of the Polyakov loop $\Phi$ in the framework of a nonlocal SU(2) chiral quark model, establishing a contact between both deconfinement order parameters at finite temperature $T$ and chemical potential $\mu$.
In our analysis, we also include the order parameter for the chiral symmetry restoration, the chiral quark condensate.

We found that $s_0$ and $\Phi$ providing us the same information for the deconfinement transition, both for the zero and finite chemical potential cases.
At zero density, the critical temperatures for both quantities coincide exactly and, at finite $\mu$ both order parameters provide evidence for the appearance of a quarkyonic phase.
\end{abstract}


\pacs{}
\maketitle


\section{Introduction}
\label{intro}

In QCD the strong interaction among quarks depends on their color charge. 
When quarks are placed in a medium, this color charge is screened due density and temperature effects~\cite{Fukushima:2010bq}. 
If the density and/or the temperature increases beyond a certain critical value, one expects that the interactions between quarks no longer confine them inside a hadron, so that they are free to travel longer distances and deconfine. 
This transition from a confined to a deconfined phase is usually referred to as the deconfinement phase transition.

A separate phase transition takes place when the realization of chiral symmetry shifts from a Nambu-Goldstone phase to a Wigner-Weyl phase. 
Based, on lattice QCD evidence~\cite{Petreczky:2012rq} one expects these two phase transitions to take place at approximately the same temperature at zero chemical potential. 
At finite density these two transitions can arise at different critical temperatures. 
The result will be a quarkyonic phase, where the chiral symmetry is restored but the quarks and gluons remains confined. 

In order to characterize the properties of these phase transitions it has been customary to study the behavior of corresponding order parameters as functions of the temperature $T$ and the baryon chemical potential $\mu$, namely the trace of the Polyakov loop (PL) $\Phi$ (deconfinement phase transition) and quark anti-quark chiral condensate $\langle\bar{\psi} \psi\rangle$ (chiral symmetry restoration), respectively. 

Another important parameter in the discussion of these phase transitions is the role that an external magnetic field may play, inducing changes in the critical temperature, in the location of the critical end point, etc~\cite{Miransky:2015ava}. 
However, in this work we will not refer to magnetic field effects, since the goal of our discussion is to compare the Polyakov loop order parameter with another QCD deconfinement parameter that has been introduced in the literature~\cite{Bochkarev:1986es} in the form of the squared energy threshold, $s_0(T)$, for the onset of perturbative QCD (PQCD) in hadronic spectral functions. 
For an actual general review see Ref.~\cite{Ayala:2016vnt}.
Around this energy, and at zero temperature, the resonance peaks in the spectrum are either no longer present or become very broad. 
The smooth hadronic spectral function thus approaches the PQCD regime.
With increasing temperature approaching the critical temperature for deconfinement, one would expect hadrons to disappear from the spectral function which should then be described entirely by PQCD.

When both $T$ and $\mu$ are nonzero, lattice QCD simulations cannot be used, because of the sign problem in the fermionic determinant. 
Therefore, one need to resort either to mathematical constructions to overcome the above limitation, or to model calculations. 

The two deconfinement order parameters mentioned before: the trace of the PL ($\Phi$) and the continuum threshold ($s_0$) can be used to realize a phenomenological description of the deconfinement transition at finite temperature and density.

The natural framework to determine $s_0$ has been that of QCD sum rules~\cite{QCDSRreview}. 
This quantum field theory framework is based on the operator product expansion (OPE) of current correlators at short distances, extended beyond perturbation theory, and on Cauchy's theorem in the complex $s$-plane. 
The latter is usually referred to as quark-hadron duality. 
Vacuum expectation values of quark and gluon field operators effectively parametrize the effects of confinement. 
An extension of this method to finite temperature was first outlined in~\cite{Bochkarev:1986es}. 
Further evidence supporting the validity of this program was provided in~\cite{Dominguez:1994re}, followed by a large number of applications~\cite{Dominguez,QCDT}. 
 
To analyze the role of the PL, we will concentrate on nonlocal Polyakov$-$Nambu$-$Jona-Lasinio (nlPNJL) models~\cite{Blaschke:2007np,Contrera:2007wu,Contrera:2009hk,Hell:2008cc,Hell:2009by,Carlomagno:2013ona}, in which quarks move in a background color field and interact through covariant nonlocal chirally symmetric four point couplings. 
These approaches, which can be considered as an improvement over the (local) PNJL model~\cite{Meisinger:1995ih,Fukushima:2003fw,Megias:2004hj,Ratti:2005jh,Roessner:2006xn,Mukherjee:2006hq,Sasaki:2006ww}, offer a common framework to study both the chiral restoration and deconfinement transitions. 
In fact, the nonlocal character of the interactions arises naturally in the context of several successful approaches to low-energy quark dynamics~\cite{Schafer:1996wv,Roberts:1994dr,Roberts:2000aa}, and leads to a momentum dependence in the quark propagator that can be made consistent~\cite{Noguera:2008cm} with lattice results~\cite{bowman,Parappilly:2005ei,Furui:2006ks}.

In view of the above mentioned points, the aim of the present work is to study the relation between both order parameters for the deconfinement transition at finite temperature and chemical potential, $\Phi$ and $s_0$, using the thermal finite energy sum rules (FESR) with inputs obtained from nlPNJL models. 


\section{Finite energy sum rules}
\label{fesr}

We begin by considering the (charged) axial-vector current correlator at $T=0$
\begin{align}
   \Pi_{\mu\nu}(q^2) &= i\int d^4x \,e^{iq\cdot x}\,
   \langle 0| T(A_\mu(x) A_\nu(0))|0 \rangle, \nonumber \\
   &= - g_{\mu\nu} \, \Pi_1(q^2) + q_\mu q_\nu \Pi_0(q^2) \;,
\label{correlator}
\end{align}
where $A_\mu(x) = :\bar{u}(x) \gamma_\mu \gamma_5 d(x):$ is the axial-vector current, $q_\mu = (\omega, \vec{q})$ is the four-momentum transfer, and the functions $\Pi_{0,1}(q^2)$ are free of kinematical singularities. 
Concentrating on the function $\Pi_0(q^2)$ and writing the OPE beyond perturbation theory in QCD \cite{QCDSRreview}, one of the two pillars of the sum rule method,  one has
\begin{equation}
\Pi_0(q^2)|_{\mbox{\tiny{QCD}}} = C_0 \, \hat{I} + \sum_{N=1} C_{2N} (q^2,\mu^2) \langle \hat{\mathcal{O}}_{2N} (\mu^2) \rangle \;, 
\label{OPE}
\end{equation}
where $\mu^2$ is a renormalization scale. 
The Wilson coefficients $C_N$ depend on the Lorentz indices and quantum numbers of the currents. 
Finally, the local gauge invariant operators ${\hat{\mathcal{O}}}_N$, are built from the quark and gluon fields in the QCD Lagrangian. The vacuum expectation values of those operators ($\hat{\mathcal{O}}_{2N} (\mu^2)$), dubbed as condensates, parameterize nonperturbative effects and have to be extracted from experimental data or model calculations. 
These operators are ordered by increasing dimensionality and the Wilson coefficients, calculable in PQCD, fall off by corresponding powers of $-q^2$. 
The unit operator above has dimension $d=0$ and $C_0 \hat{I}$ stands for the purely perturbative contribution. 
Hence, this OPE factorizes short distance physics, encapsulated in the Wilson coefficients, and long distance effects parametrized by the vacuum condensates. 

The second pillar of the QCD sum rules technique is Cauchy's theorem in the complex squared energy $s$-plane
\begin{align}
\frac{1}{\pi}\int_{0}^{s_0} ds\ f(s)\ \mbox{Im}\  \Pi_0 (s)|_{\mbox{\tiny{HAD}}} = \nonumber \\
   -\frac{1}{2\pi i}\oint_{C(|s_0|)}ds\ f(s)\ & \Pi_0 (s)|_{\mbox{\tiny{QCD}}} \;,
\label{disprel}
\end{align}
where $f(s)$ is an arbitrary analytic function, and the radius of the circle $s_0$ is large enough for QCD and the OPE to be used on the circle. 
The integral along the real $s$-axis involves the hadronic spectral function. 
This equation is the mathematical statement of what is usually referred to as quark-hadron duality. 
Using the OPE, Eq.(\ref{OPE}), and an integration kernel $f(s) = s^N \; (N=1,2,\cdots)$ one obtains the  FESR
\begin{align}
(-)^{N-1} C_{2N} \langle {\mathcal{\hat{O}}}_{2N}\rangle = 4 \pi^2 \int_0^{s_0} ds\, s^{N-1} \,\frac{1}{\pi} {\mbox{Im}} \Pi_0(s)|_{\mbox{\tiny{HAD}}}
\nonumber \\
- \frac{s_0^N}{N} \left[1+{\mathcal{O}}(\alpha_s)\right] \;\; (N=1,2,\cdots) \;.
\label{FESR}
\end{align}

For $N=1$, the dimension $d=2$ term in the OPE does not involve any condensate, as it is not possible to construct a gauge invariant operator of such a dimension from the quark and gluon fields. 
There is no evidence for such a term (at $T=0$) from FESR analyses of experimental data on $e^+ e^-$ annihilation and $\tau$ decays into hadrons \cite{Dominguez:1999xa, Dominguez:2006ct}. 
At high temperatures, though, there seems to be evidence for some $d=2$ term \cite{Megias:2009ar}.
However, the analysis to be reported here is performed at lower values of $T$, so that we can safely ignore this contribution in the sequel. 

The dimension $d=4$ term, a renormalization group invariant quantity, is given by 
\begin{equation}
C_4 \langle \hat{\mathcal{O}}_{4}  \rangle = 
\frac{\pi}{6} \langle \alpha_s G^2\rangle + 2 \pi^2 (m_u + m_d) \langle\bar{q} q \rangle , 
\label{C4}
\end{equation}
The leading power correction of dimension $d=6$ is the four-quark condensate, which in the vacuum saturation approximation~\cite{QCDSRreview} becomes
\begin{equation}
C_6 \langle \hat{\mathcal{O}}_{6}  \rangle = \frac{896}{81} \,\pi^3 \, \alpha_s \,|\langle \bar{q} q \rangle|^2\;, 
\label{C6}
\end{equation}
which has a very mild dependence on the renormalization scale. 
This  approximation has no solid theoretical justification, other than its simplicity. 
Hence, there is no reliable way of estimating corrections, which in fact appear to be rather large from comparisons between Eq. (\ref{C6}) and direct determinations from data~\cite{Dominguez:2006ct}. 

The extension of this program to finite temperature is fairly straightforward~\cite{Bochkarev:1986es,Dominguez:1994re,Ayala:2011vs}, with the Wilson coefficients in the OPE, Eq.(\ref{OPE}), remaining independent of $T$ at leading order in $\alpha_s$, and the condensates developing a temperature dependence. 
Radiative corrections in QCD  involve now an additional scale, i.e. the temperature, so that $\alpha_s \equiv \alpha_s(\mu^2,T)$. 
This problem has not yet been  solved successfully. 
Nevertheless, from the size of radiative corrections at $T=0$ one does not expect any major loss of accuracy in results from thermal FESR to leading order in PQCD, as long as the temperature is not too high, say $T \lesssim 200 \, {\mbox {MeV}}$. 
Essentially all applications of  FESR at $T \neq 0$ have been done at leading order in PQCD, thus implying a systematic uncertainty at the level of 10 \%. 

In the static limit ($\vec{q} \rightarrow 0$), to leading order in PQCD, and for $T\neq 0$ and $\mu \neq 0$  the  function $\Pi_0(q^2)|_{\mbox{\tiny{QCD}}}$ in Eq.(\ref{correlator}) becomes $\Pi_0(\omega^2, T, \mu)|_{\mbox{\tiny{QCD}}}$; to simplify the notation we shall omit the $T$ and $\mu$ dependence in the sequel. 
A straightforward calculation of the spectral function in perturbative QCD, at finite temperature and finite density gives
\begin{align}
   \frac{1}{\pi} {\mbox{Im}}\Pi_0(s)|_{\mbox{\tiny{PQCD}}}
   =
   \frac{1}{4\pi^2}\left[1-\tilde{n}_+\left(\frac{\sqrt{s}}{2}\right) 
   -\tilde{n}_-\left(\frac{\sqrt{s}}{2}\right)\right] \nonumber \\
   -\frac{2}{\pi^2} \;T^2 \;\delta (s)\; \left[
   {\mbox{Li}}_2(-e^{\mu/T})
   + {\mbox{Li}}_2(-e^{-\mu/T})\right],
\label{pertQCD}
\end{align}
where ${\mbox{Li}}_2(x)$ is the dilogarithm function, $s=\omega^2$, and
\begin{equation}
   \tilde{n}_\pm(x)=\frac{1}{e^{(x\mp \mu)/T}+1}
\label{F-D}
\end{equation}
are the Fermi-Dirac thermal distributions for particles and antiparticles,
respectively. 

In the hadronic sector we assume pion-pole dominance of the hadronic spectral function, i.e. the continuum threshold $s_0$ to lie below the first radial excitation with mass $M_{\pi_1} \simeq 1 300\;{\mbox{MeV}}$. 
This is a very good approximation at finite $T$, as we expect $s_0$ to be monotonically decreasing with increasing temperature. 
In this case, 
\begin{equation}
   \frac{1}{\pi}{\mbox{Im}}\Pi_0 (s)|_{\mbox{\tiny{HAD}}}
   = 2 \; f_\pi^2(T,\mu_B)\; \delta (s-m_\pi^2),
\label{HAD}
\end{equation}
where $f_\pi(T,\mu_B)$ is the pion decay constant at finite $T$ and $\mu$, with $f_\pi(0,0) =92.21 \pm 0.14 \;{\mbox{MeV}}$ \cite{Agashe:2014kda}. 
Notice we will not include in our spectral function the first part of $a_1$ resonance obtained from the $\tau$-decay data~\cite{Dominguez:2012bs}, since still there is no counterpart in the SU(2) nlPNJL model for the description of the hadronic vector resonance. 
A zero temperature analysis has been done for the vector case in Ref.~\cite{Villafane:2016ukb}.

Turning to the FESR, Eq.(\ref{FESR}), with $N=1$ and no dimension $d=2$ condensate, and using Eqs.(\ref{pertQCD}) and (\ref{HAD}) one finds
\begin{align}
\frac{}{}\int_0^{s_0(T,\mu)}  ds \, \left[1 - \tilde{n}_+\left(\frac{\sqrt{s}}{2}\right) 
   -\tilde{n}_-\left(\frac{\sqrt{s}}{2}\right)\right] = \nonumber\\
   8 \pi^2 f_\pi^2(T,\mu) +  8 T^2 \left[{\mbox{Li}}_2(-e^{\mu/T})
   + {\mbox{Li}}_2(-e^{-\mu/T})\right] .
\label{FESRTMU}
\end{align}
This is a transcendental equation determining $s_0(T,\mu)$ in terms of $f_\pi(T,\mu)$.  

For completeness, the other two thermal FESR at zero chemical potential are given by \cite{Dominguez:2012bs},
\begin{align}
- C_{4}\langle {\mathcal{\hat{O}}}_{4}\rangle(T) = 4 \pi^2 \int_0^{s_0(T)} ds\, s \frac{1}{\pi} {\mbox{Im}}\, \Pi_0(s)|_{\mbox{\tiny{HAD}}}
\nonumber \\
-  \int_0^{s_0(T)}ds \, s \left[1 - 2  n_F\left(\frac{\sqrt{s}}{2 T}\right)\right] ,
\label{FESRT2}
\end{align}
\begin{align}
 C_{6}\langle {\mathcal{\hat{O}}}_{6}\rangle(T) = 4 \pi^2 \int_0^{s_0(T)} ds\, s^2 \frac{1}{\pi} {\mbox{Im}}\, \Pi_0(s)|_{\mbox{\tiny{HAD}}}
\nonumber \\
-  \int_0^{s_0(T)}ds \; s^2 \left[1 - 2  n_F\left(\frac{\sqrt{s}}{2 T}\right)\right] \;,
\label{FESRT3}
\end{align}
where $n_F(x)=1/(1+e^x)$ is the Fermi thermal function.


\section{Thermodynamics at finite density in the PNJL model}
\label{thermo}

We consider a nonlocal SU(2) chiral quark model that includes quark couplings to the color gauge fields. 
The corresponding Euclidean effective action is given by~\cite{Contrera:2010kz,Pagura:2011rt}
\begin{align}
 S_{E} = \int d^{4}x\ \bigg\{
\bar{\psi}(x)\left( -i\gamma_{\mu}D_{\mu}
+\hat{m}\right)  \psi(x) - \nonumber \\
\left. \frac{G_{S}}{2} \Big[ j_{a}(x)j_{a}(x)- j_{P}%
(x)j_{P}(x)\Big]+ \ {\cal U}\,(\Phi[A(x)]) \right\rbrace \ , 
\label{action}%
\end{align}
where $\psi$ is the $N_{f}=2$ fermion doublet $\psi\equiv(u,d)^T$, and $\hat{m}={\rm diag}(m_{u},m_{d})$ is the current quark mass matrix. In what follows we consider isospin symmetry, $m_{u}=m_{d}=m$. 
The fermion kinetic term in Eq.~(\ref{action}) includes a covariant derivative $D_\mu\equiv\partial_\mu - iA_\mu$, where $A_\mu$ are color gauge fields. 
The nonlocal currents $j_{a}(x),j_{P}(x)$ are given by
\begin{align}
j_{a}(x)  &  =\int d^{4}z\ {\cal G}(z)\ \bar{\psi}\left(  x+\frac{z}{2}\right)
\ \Gamma_{a}\ \psi\left(  x-\frac{z}{2}\right)  \ ,\nonumber\\
j_{P}(x)  &  =\int d^{4}z\ {\cal F}(z)\ \bar{\psi}\left(  x+\frac{z}{2}\right)
\ \frac{i {\overleftrightarrow{\rlap/\partial}}}{2\ \kappa_{p}}
\ \psi\left(  x-\frac{z}{2}\right)\ ,
\label{currents}%
\end{align}
where,
$\Gamma_{a}=(\leavevmode\hbox{\small1\kern-3.8pt\normalsize1},i\gamma
_{5}\vec{\tau})$ and $u(x^{\prime}){\overleftrightarrow{\partial}%
}v(x)=u(x^{\prime})\partial_{x}v(x)-\partial_{x^{\prime}}u(x^{\prime})v(x)$.
The functions ${\cal G}(z)$ and ${\cal F}(z)$ in Eq.~(\ref{currents}) are nonlocal covariant form factors characterizing the corresponding interactions. 

Notice that the four currents $j_a(x)$ require a common form factor ${\cal G}(z)$ in order to guarantee chiral invariance, while the coupling $j_{P}(x)j_{P}(x)$ is self-invariant under chiral transformations. 
The scalar-isoscalar component of the $j_{a}(x)$ current will generate a momentum dependent quark mass in the quark propagator, while the ``momentum'' current $j_{P}(x)$ will be responsible for a momentum dependent quark wave function renormalization (WFR)~\cite{Noguera:2008cm,Contrera:2010kz,Pagura:2011rt}, if is not included then the mass parameter in the quark propagator cannot be compare with lattice results.

Now we perform a bosonization of the theory, introducing bosonic fields $\sigma_{1,2}(x)$ and $\pi_a(x)$, and integrating out the quark fields. 
Details of this procedure can be found e.g.\ in Ref.~\cite{Noguera:2008cm}. 

In order to analyze the properties of meson fields it is necessary to go beyond the mean field approximation, considering quadratic fluctuations in the Euclidean action:
\begin{align}
\label{spiketa}
S_E^{\rm quad} &=& \dfrac{1}{2} \int \frac{d^4 p}{(2\pi)^4} \sum_{M}\  r_M\
G_M(p^2)\  \phi_M(p)\, \bar\phi_M(-p) \ ,
\end{align}
where meson fluctuations $\delta\sigma_a$, $\delta\pi_a$  have been translated to a charged basis $\phi_M$, being $M$ the scalar and pseudoscalar mesons ($\sigma,\pi^0$, $\pi^\pm$) plus the $\sigma_2$ field, and $G_M$ are the inverse dressed propagators.
The coefficient $r_M$ is 1 for charge eigenstates $M=\sigma_i,\pi^0$, and 2 for $M=\pi^+$. 
Meson masses are then given by the equations
\begin{equation}
G_M(-m_M^2)\ =\ 0 \ ,
\end{equation}
where the full expressions for the one-loop functions $G_M(q)$ can be found in Ref.~\cite{Noguera:2008cm,Carlomagno:2013ona}. 
In addition, physical states have to be normalized through
\begin{equation}
\tilde{\phi}_M(p)=Z_M^{-1/2}\ \phi_M(p)\ ,
\end{equation}
where
\begin{equation}
\label{zr}
Z_M^{-1}=\frac{dG_M(p)}{dp^2}\bigg\vert_{p^2=-m_M^2} \ .
\end{equation}

At finite temperature, the meson masses are obtained by solving $G_P (- m_P^2,  0) = 0$. 
The mass values determined by these equations are the spatial ``screening-masses'' corresponding to the zeroth Matsubara mode, and their inverses describe the persistence lengths of these modes at equilibrium with the heat bath~\cite{Contrera:2009hk}.

At zero temperature, one can also calculate the weak decay constants of pseudoscalar mesons. 
These are given by the matrix elements of the axial currents $ A_\mu^a$ between the vacuum and the physical meson states,
\begin{equation}
\label{fpiab}
\imath f_{ab}(p^2) \; p_\mu=\langle 0 \vert A_\mu^a(0) \vert \delta\pi_b(p)
\rangle\ .
\end{equation}
The matrix elements can be calculated from the expansion of the Euclidean effective action in the presence of external axial currents,
\begin{equation}
\label{der}
\langle 0 \vert A_\mu^a(0) \vert \delta\pi_b(p) \rangle = \frac{\delta^2
S_E}{\delta A_\mu^a \delta \pi_b(p)}\bigg\vert_{A_\mu^a=\delta\pi_b=0} \ ,
\end{equation}
Performing the derivative of the resulting expressions with respect to the renormalized meson fields, we can finally identify the corresponding pion weak decay constant~\cite{Noguera:2008cm,Carlomagno:2013ona}
\begin{equation}
f_{\pi}=\frac{m_{c}\; Z^{-1/2}_\pi}{m_{\pi}^{2}}\; F_{0}(-m_{\pi}^{2})\ .
\label{fpi}%
\end{equation}
with
\begin{align}
F_{0}(p^{2})=8\, N_{c}\int\frac{d^{4}q}{(2\pi)^{4}}\ g(q)\;\frac
{Z(q^{+})Z(q^{-})}{D(q^{+})D(q^{-})}\times \nonumber \\
\left[q^{+}\cdot q^{-}+M(q^{+})M(q^{-})\right]
\label{fpi2}
\end{align}
where $q^{\pm}=q\pm p/2\,$ and $D(q)=q^{2}+M^{2}(q)$, with $M(p)$ and $Z(p)$ defined as
\begin{align}
M(p) &=  Z(p) \left[m_q + \bar\sigma_1 \ g(p) \right] \ , \nonumber \\
Z(p) &=  \left[ 1 - \bar\sigma_2 \ f(p) \right]^{-1}\ .
\label{mz}
\end{align}
here $g(p)$ and $f(p)$ are the Fourier transforms of the form factors in Eq.~(\ref{currents}).

\vspace*{0.5cm}

Since we are interested in the deconfinement and chiral restoration critical temperatures, we extend the bosonized effective action to finite temperature $T$ and chemical potential $\mu$. 
This will be done using the standard imaginary time formalism. 
Concerning the gauge fields $A_\mu$, we assume that quarks move on a constant background field $\phi = A_4 = i A_0 = i g\,\delta_{\mu 0}\, G^\mu_a \lambda^a/2$, where $G^\mu_a$ are SU(3) color gauge fields. 
Then the traced Polyakov loop, which in the infinite quark mass limit can be taken as an order parameter of confinement, is given by $\Phi=\frac{1}{3} {\rm Tr}\, \exp( i \phi/T)$. 
For the light quark sector the trace of the Polyakov loop turn out to be an approximate order parameter in the same way the chiral quark condensate is an approximate order parameter for the chiral symmetry restoration outside the chiral limit.

We work in the so-called Polyakov gauge~\cite{Diakonov:2004kc}, where the matrix $\phi$ is given a diagonal representation $\phi = \phi_3 \lambda_3 + \phi_8 \lambda_8$. 
This leaves only two independent variables, $\phi_3$ and $\phi_8$. 
Owing to the charge conjugation properties of the QCD Lagrangian, the expectation values $\langle \Phi \rangle$ and $\langle \Phi^* \rangle$ of the conjugate Polyakov loop fields must be real quantities~\cite{Dumitru:2005ng,Roessner:2006xn}. 
This means $ \Phi = \Phi^*$ for the mean field configurations that satisfy the gap equations. 
With the constraint of $\phi_3$ and $\phi_8$ being real: $\phi_8=0$, leaving only $\phi_3$ as an independent variable, and therefore $\Phi = [ 2 \cos(\phi_3/T) + 1 ]/3$.

Thus, in the mean field approximation (MFA), and following the same prescriptions as in previous works, see e.g.~Refs.~\cite{GomezDumm:2001fz,GomezDumm:2004sr}, the thermodynamical potential $\Omega^{\rm MFA}$ at finite temperature $T$ and chemical potential $\mu$ is given by
\begin{equation}
\label{omegareg}
\Omega^{\rm MFA} \ = \ \Omega^{\rm reg} + \Omega^{\rm free} +
\mathcal{U}(\Phi,T) + \Omega_0 \ ,
\end{equation}
where
\begin{widetext}
\begin{align}
\Omega^{\rm reg} &=  \,- \,4 T \sum_{c=r,g,b} \ \sum_{n=-\infty}^{\infty}
\int \frac{d^3\vec p}{(2\pi)^3} \ \log \left[ \frac{ (\rho_{n,
\vec{p}}^c)^2 + M^2(\rho_{n,\vec{p}}^c)}{Z^2(\rho_{n, \vec{p}}^c)}\right]+
\frac{\bar\sigma_1^2 + \kappa_p^2\; \bar\sigma_2^2}{2\,G_S} \ , \nonumber \\
\Omega^{\rm free} \ &= \ -4 T \int \frac{d^3 \vec{p}}{(2\pi)^3}\;
\sum_{c=r,g,b} \ \sum_{s=\pm 1}\mbox{Re}\;
\ln\left[ 1 + \exp\left(-\;\frac{\epsilon_p + i s \phi_c}{T}
\right)\right]
\ ,
\label{granp}
\end{align}
\end{widetext}
here $\bar\sigma_{1,2}$ are the mean field values of the scalar fields. We have also defined
\begin{equation}
\Big({\rho_{n,\vec{p}}^c} \Big)^2 =
\Big[ (2 n +1 )\pi  T + \phi_c - \imath \mu \Big]^2 + {\vec{p}}\ \! ^2 \ , 
\end{equation}
the sums over color indices run over $c=r,g,b$, with the color background fields components being $\phi_r = - \phi_g = \phi_3$, $\phi_b = 0$, and $\epsilon_p = \sqrt{\vec{p}^{\;2}+m^2}\;$. 
The term $\Omega^{\rm reg}$ is the regularized expression with the thermodynamical potential of a free fermion gas, and finally the last term in Eq.~(\ref{omegareg}) is just a constant fixed by the condition that $\Omega^{\rm MFA}$ vanishes at $T=\mu=0$.

The effective gauge field self-interactions are given by the Polyakov loop potential $\mathcal{U}(\Phi,T)$. At finite temperature $T$, it is usual to take for this potential a functional form based on properties of pure gauge QCD. 
One possible Ansatz is that based on the logarithmic expression of the Haar measure associated with the SU(3) color group integration. 
The corresponding potential is given by~\cite{Roessner:2006xn}
\begin{align}
\frac{{\cal{U}}_{\rm log}(\Phi ,T)}{T^4} =\ -\,\frac{1}{2}\, a(T)\,\Phi^2 \;+\nonumber \\ 
\;b(T)\, \log\left(1 - 6\, \Phi^2 + 8\, \Phi^3
- 3\, \Phi^4 \right) \ ,
\label{ulog}
\end{align}
where
\begin{align}
a(T) &= a_0 +a_1 \left(\dfrac{T_0}{T}\right) + a_2\left(\dfrac{T_0}{T}\right)^2 \ , \nonumber\\
b(T) &= b_3\left(\dfrac{T_0}{T}\right)^3 \ .
\label{log}
\end{align}
The parameters can be fitted to pure gauge lattice QCD data to properly reproduce the corresponding equation of state and the Polyakov loop behavior~\cite{Roessner:2006xn}. 
The values of $a_i$ and $b_i$ are constrained by the condition of reaching the Stefan-Boltzmann limit at $T \rightarrow \infty$ and by imposing the presence of a first-order phase transition at $T_0$, which is a further parameter of the model. 
At the critical temperature, the Polyakov loop potential develops a second degenerate minimum giving raise to a first order phase transition.

In the absence of dynamical quarks, from lattice calculations one expects a deconfinement temperature $T_0 = 270$~MeV. 
However, it has been argued that in the presence of light dynamical quarks this temperature scale should be adequately reduced to about 210 and 190~MeV for the case of two and three flavors, respectively, with an uncertainty of about 30 MeV~\cite{Schaefer:2007pw}. In this work we will use $T_0 = 208$~MeV.

Besides the logarithmic function in Eq.~(\ref{ulog}), a widely used potential is that given by a polynomial function based on a Ginzburg-Landau Ansatz~\cite{Ratti:2005jh,Scavenius:2002ru}:
\begin{align}
\frac{{\cal{U}}_{\rm poly}(\Phi ,T)}{T ^4} \ = \ -\,\frac{b_2(T)}{2}\, \Phi^2
-\,\frac{b_3}{3}\, \Phi^3 +\,\frac{b_4}{4}\, \Phi^4 \ ,
\label{upoly}
\end{align}
where
\begin{align}
b_2(T) = a_0 +a_1 \left(\dfrac{T_0}{T}\right) + a_2\left(\dfrac{T_0}{T}\right)^2
+ a_3\left(\dfrac{T_0}{T}\right)^3\ .
\label{pol}
\end{align}
Once again, the parameters can be fitted to pure gauge lattice QCD results to reproduce the corresponding equation of state and Polyakov loop behavior (numerical values can be found in Ref.~\cite{Ratti:2005jh}).

Given the full form of the thermodynamical potential, the mean field values $\bar\sigma_{1,2}$ and $\phi_{3}$ can be obtained as solutions of the coupled set of gap equations
\begin{equation}
\frac{\partial \Omega^{\rm MFA}_{\rm reg}}
{\left(\partial\sigma_{1},\partial\sigma_{2}, \partial\phi_3\right)}\ = \ 0 \ .
\label{fullgeq}
\end{equation}

In order to fully specify the model under consideration, we proceed to fix the model parameters as well as the nonlocal form factors $g(q)$ and $f(q)$. We consider here Gaussian functions 
\begin{align}
g(q) &= \mbox{exp}\left(-q^{2}/\Lambda_{0}^{2}\right) \ , \nonumber \\
f(q) &= \mbox{exp}\left(-q^{2}/\Lambda_{1}^{2}\right)\ ,
\label{regulators}
\end{align}
which guarantee a fast ultraviolet convergence of the loop integrals. 
The values of the five free parameters can be found in~\cite{Pagura:2011rt}.

Once the mean field values are obtained, the behavior of other relevant quantities as functions of the temperature and chemical potential can be determined. 
We concentrate, in particular, on the chiral quark condensate $\langle\bar{q}q\rangle = \partial\Omega^{\rm MFA}_{\rm reg}/\partial m$ and the traced Polyakov loop $\Phi$, which will be taken as order parameters for the chiral restoration and deconfinement transitions, respectively. 
The associated susceptibilities will be defined as $\chi_{\rm ch}  = \partial\,\langle\bar qq\rangle/\partial m$ and $\chi_{\rm PL} = d \Phi / d T$. 


\section{Results}
\label{results}

In order to determine the relation between both order parameters for the deconfinement transition, namely the perturbative QCD threshold $s_0$ and the trace of the Polyakov loop $\Phi$ as functions of the temperature and chemical potential we begin our analysis studying the finite energy sum rules at zero density. 
In this scenario,  when $\mu=0$, the Eq.~(\ref{FESRTMU}) becomes
\begin{align}
8  \pi^2 f^2_\pi(T) &=& \frac{4}{3}  \pi^2  T^2  + \int_0^{s_0(T)}ds \,\left[1 - 2\, n_F \left(\frac{\sqrt{s}}{2 T} \right) \right] \;, \label{FESRT1}
\end{align}
where the pion decay constant at finite temperature and/or chemical potential is calculated using the Eq.~(\ref{fpi}) with Eq.~(\ref{fpi2}) as
\begin{align}
F_{0}(p^{2})=8\, T \sum_{c,n} \int\frac{d^{3}\vec{q}}{(2\pi)^{4}}\ g({\rho_{n,\vec{q}}^c})\; 
\frac{Z({\rho_{n,\vec{q}}^c}^{+})Z({\rho_{n,\vec{q}}^c}^{-})}{D({\rho_{n,\vec{q}}^c}^{+})D({\rho_{n,\vec{q}}^c}^{-})}\ \times \nonumber \\
\left[  {\rho_{n,\vec{q}}^c}^{+}\cdot {\rho_{n,\vec{q}}^c}^{-}+M({\rho_{n,\vec{q}}^c}^{+}%
)M({\rho_{n,\vec{q}}^c}^{-})\right]
\label{fpi3}
\end{align}
where ${\rho_{n,\vec{q}}^c}^{\pm}={\rho_{n,\vec{q}}^c} \pm p/2\,$.

\begin{figure}[h]
\includegraphics[width=0.45\textwidth]{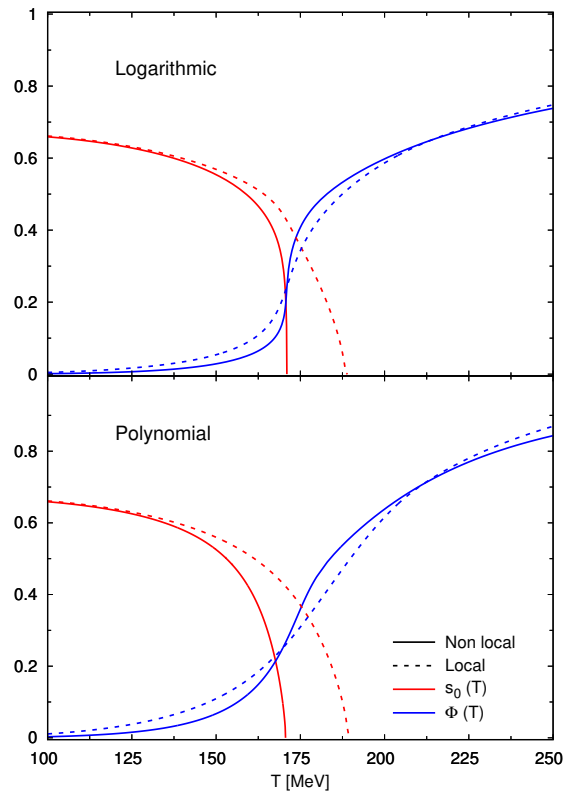}
\caption{\label{fig:lvsnl} Continuum threshold (red line) and trace of the Polyakov loop (blue line) as a function of the temperature for nonlocal (solid line) and local PNJL model (dashed line) at zero chemical potential for logarithmic (upper panel) and polynomial (lower panel) effective potentials.}
\end{figure}

It is known that in local versions of the PNJL model, at zero chemical potential, the restoration of the chiral symmetry and the deconfinement transition take place at different temperatures (see e.g.~Refs.~\cite{Fu:2007xc,Costa:2008dp}), usually separated by approximate $20$~MeV. 
Therefore, it is interesting to analyze the results obtained in a nonlocal and in a local PNJL model, the latter one parametrized according to~\cite{Ratti:2005jh}. 
In Fig.~\ref{fig:lvsnl} we plot the continuum threshold and the trace of the PL for the nonlocal (local) PNJL model in solid (dashed) line, for the logarithmic and polynomial effective potentials. 
As we expected from previous results, in the local version both transitions do not occur simultaneously. 
In this scenario, the PQCD threshold vanishes at a critical temperature, $T_c^{s_0}$, located between the chiral critical temperature $T_c^{\chi}$ and the PL deconfinement temperature $T_c^{\Phi}$ (obtained through the corresponding susceptibilities) and hence, although it is not possible to conclude a direct relation between $s_0$ and $\Phi$, the continuum threshold, in any case, vanishes before the restoration of the chiral symmetry, in agreement with general arguments~\cite{Bochkarev:1986es}.

In the case of the nonlocal PNJL model, for both effective potentials, $s_0$ and $\Phi$ have a similar critical temperature for the deconfinement transition of approximate $T_c \sim 170$ MeV.
These temperatures are summarized in Table~\ref{TableI:tes_c}.
\begin{table}
\begin{tabular}{c c c c c}
\hline 
 & \multicolumn{2}{c}{Logarithmic} & \multicolumn{2}{c}{Polynomial} \\
\hline 
 & Non local & Local & Non local & Local \\
\hline
$T_c^{\chi}$ [MeV] & 171 & 205 & 176 & 201 \\
$T_c^{\Phi}$ [MeV] & 171 & 171 & 174 & 183 \\
$T_c^{s_0}$ [MeV]  & 171 & 189 & 170 & 190 \\
\hline
\end{tabular}
\caption{Chiral critical temperatures $T_c^{\chi}$, deconfinement temperatures $T_c^{\Phi}$ and $T_c^{s_0}$ for the local and nonlocal PNJL model with logarithmic and polynomial effective potentials.}
\label{TableI:tes_c}
\end{table}

The value obtained at zero temperature for the continuum threshold, $s_0 \sim 670$, MeV is rather small but in a good agreement with other calculations in sum rules using as input LQCD results.
The main reason for this lower value is the pion pole approximation for the spectral function. When additional information is incorporated, for instance the $a_1$ resonance, the value of $s_0(T=0)$ increases substantially~\cite{Dominguez:2012bs}.

\begin{figure}[h]
\includegraphics[width=0.45\textwidth]{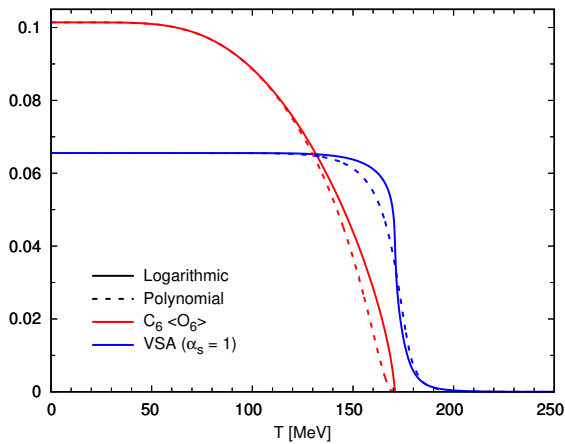}
\caption{\label{fig:qqqq} Four-quark condensate in the vacuum saturation approximation with $\alpha_s=1$~\cite{Deur:2016tte} (blue line) and $C_6 \langle \hat{\mathcal{O}}_{6} \rangle$ (red line) for the logarithmic (polynomial) Polyakov effective potential in solid (dashed) line, at zero density as a function of the temperature.}
\end{figure}

Just for completeness and, in addition to the main goal of this article, from the higher order FESR, Eqs.~(\ref{FESRT2}) and (\ref{FESRT3}), we can  estimate the gluon condensate and the four-quark condensate. 
The former shows the expected behavior with a finite value at zero temperature. It decreases monotonically as function of temperature, vanishing at $T \sim 170$~MeV. 
The four quark condensate, plotted in Fig.~\ref{fig:qqqq}, was compared, according to the vacuum saturation approximation (VSA), with the squared of the chiral quark condensate obtained within the $SU(2)$ nlPNJL model. 
If we assume that the previous approximation is exact, from Eqs.~(\ref{C6}) and (\ref{FESRT3}), at zero temperature and in the chiral limit, we obtain that $\alpha_s = \dfrac{108\ \pi^3}{7}\dfrac{f_\pi^6}{|\langle \bar{q}q \rangle|^2} \simeq 1.6$ (a very similar result is obtained outside the chiral limit), meaning that the VSA underestimate $C_6 \langle \hat{\mathcal{O}}_{6} \rangle$.
This value is considerably higher than recent estimations of the strong coupling at low energies, based on completely different approaches~\cite{Pich:2016bdg,Deur:2016tte}. The first one relies on a recent analysis of the ALEPH data for the $\tau$ decay, whereas the second one corresponds to a general recent review including different perspectives.

From Fig.~\ref{fig:qqqq}, we see that for both Polyakov effective potentials, the VSA is about $40 \%$ less than the four-quark condensate obtained from FESR at zero temperature, in qualitatively agreement with estimates, based on $K^0-\bar{K}^0$ mixing~\cite{Chetyrkin:1988yr}. 

From lattice QCD calculations, at zero chemical potential, the chiral symmetry restoration and the deconfinement transition take place at the same critical temperature. This behavior was verified in nlPNJL models~\cite{Contrera:2010kz,Carlomagno:2013ona} and also obtained by finite energy sum rules~\cite{Ayala:2011vs}. 
The next natural step is to extend our analysis to a finite density scenario, to identify the relation between $s_0(T,\mu)$ and $\Phi(T,\mu)$.

\begin{figure}[h]
\includegraphics[width=0.45\textwidth]{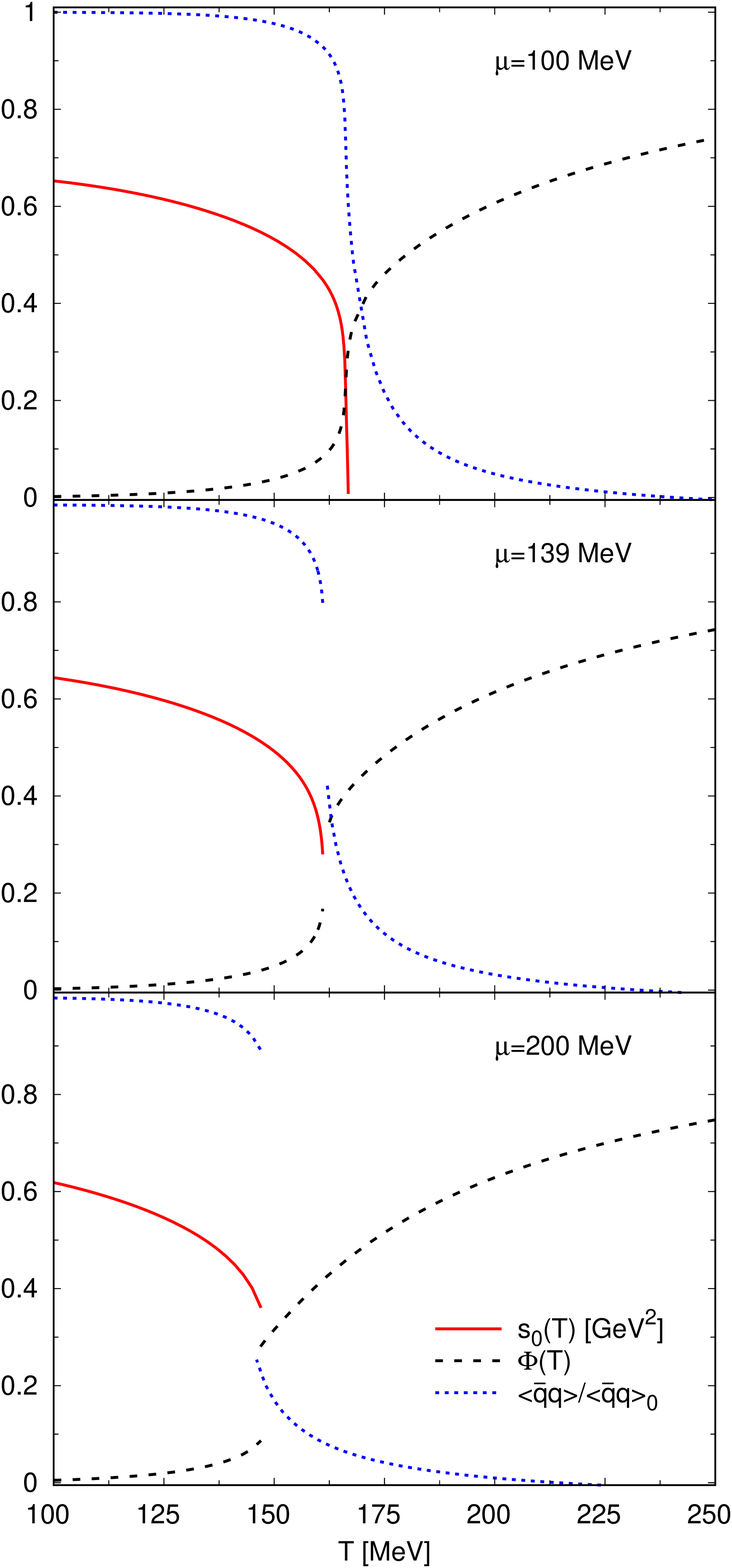}
\caption{\label{fig:munocero} Continuum threshold (solid red line), trace of the Polyakov loop (black dashed lined) and the normalized quark condensate (blue dotted line) as a function of the temperature at a constant density for the logarithmic effective potential.}
\end{figure}

In Fig.~\ref{fig:munocero} we plot, for the logarithmic Polyakov effective potential, the normalized quark condensate $\langle\bar qq\rangle/\langle\bar qq\rangle_0$, the trace of the PL $\Phi$ and the continuum threshold $s_0$ as functions of the temperature for three different values of chemical potential. 
In the middle panel we choose $\mu=139$~MeV, which correspond to the critical end point chemical potential $\mu_{CEP}$. For values of $\mu$ smaller than $\mu_{CEP}$, the chiral restoration arises via a crossover transition. Beyond this critical density, a first order phase transition occurs. 
This value, together with the critical temperature $T_{CEP} = 161$~MeV determines the coordinates of the critical end point.

All the results presented here were obtained by Gaussian regulators (see Eq.~(\ref{regulators})). 
Nevertheless, similar outcomes would be obtained if other form factors would have been employed.
For instance, a lattice inspired dependence (Lorentzian regulator)~\cite{Carlomagno:2013ona} or we may also neglect the momentum current, this means no WFR effects~\cite{Contrera:2009hk}.
It turns out that the chiral and deconfinement critical temperatures get a minor dependence on the explicit shape used to parameterize the form factors~\cite{Pagura:2011rt,Carlomagno:2015hea}.

In the upper panel of Fig.~\ref{fig:munocero}, where $\mu=100$ MeV, we see that the chiral and deconfinement transitions are crossovers occurring at the same critical temperature. 
The peak of the Polyakov susceptibility and the point where the continuum threshold vanishes occur at approximate the same temperature $T_c \sim 166$ MeV.

When $\mu$ becomes equal or higher than $\mu=139$~MeV, the order parameter for the chiral symmetry restoration has a discontinuity signaling a first order phase transition. 
These gap in the quark condensate induces also a jump in the trace of the PL (see middle and lower panels in Fig.~\ref{fig:munocero}). The value of $\Phi$ at the discontinuity indicates that at this temperature the system remains confined but in a chiral symmetry restored state. This region is usually referred as the quarkyonic phase~\cite{McLerran:2007qj,McLerran:2008ua}.

At bigger densities than the critical end point chemical potential, the thermal equation has not solution beyond the critical temperature. 
The term proportional to the dilogarithm becomes too negative and therefore Eq.~\ref{FESRTMU} can not be satisfied. 
The continuum threshold stops with a finite value at the chiral critical temperature (see middle and lower panels in Fig.~\ref{fig:munocero}). 
We see in this way, that the Polyakov loop and the continuum threshold provide the same information. When the chiral symmetry is restored, $s_0$ and $\Phi$ show that we are still in a confined phase. This characterize the occurrence of a quarkyonic phase.


\section{Summary and conclusions}
\label{finale}

In this article we discuss if the behavior of two vastly used order parameters for the deconfinement transition: the continuum threshold and the trace of the Polyakov loop, provide us with the same physical insight. 

To accomplish this analysis, we use finite energy sum rules for the axial-vector current correlator. 
In this framework, one can define the continuum threshold as the energy where the resonance peaks in the spectrum become very broad. 

On the other side, the Polyakov loop is a thermal Wilson loop, gauge-invariant under the center of the color group and is expected to vanish in the confined phase and being different from zero in the deconfined phase.

The idea was to carry on the FESR program saturating the spectral function with the pion pole approximation.
The input parameters we used in the spectral function, namely the pion mass, the pion decay constant and the chiral quark condensate, were obtained from a nonlocal SU(2) Polyakov-NJL model with Gaussian form factors.
In this way we establish the connection between both approaches.

At zero density, we compare the trace of the Polyakov loop and the continuum threshold for the local and the nonlocal version of a PNJL model. 
We determine, for the nlPNJL model, that the continuum threshold vanishes at the same temperature where the Polyakov susceptibility has its maximum value. 
In the case of the local PNJL, $s_0$ becomes zero between the critical temperature for the deconfinement transition, according to the Polyakov loop analysis, and the chiral restoration temperature. The fact that both deconfinement temperatures are smaller than the chiral critical temperature is in agreement with other analysis.

At finite chemical potential, we find that for both deconfinement parameters, beyond the critical end point chemical potential, the system remains in its confined phase even when the chiral symmetry is restored. 
This is an evidence for the appearance of a quarkyonic phase. 
 
We may conclude saying that our analysis gives strong support to the idea that both deconfinement parameters, in fact, provide the same kind of physical information.


\section*{Acknowledgements}

This work has been partially funded by CONICET (Argentina) under Grant No.\ PIP 449; by the National University of La Plata (Argentina), Project No.\ X718; by FONDECYT (Chile), under grants No. 1130056, 1150171 and 1150847; and by Proyecto Basal (Chile) FB 0821. 



\end{document}